\def\mincir{\ \raise-2.truept\hbox{\rlap{\hbox{$\sim$}}\raise5.truept
    \hbox{$<$}\ }}
\def\bd30{BD+30\degr3639}

\def\kms{km\,s$^{-1}$}

\def\nii{[N{\sc ii}]}

\def\oiii{[O{\sc iii}]}
\def\ha{H$\alpha$}
\def\hanii{H$\alpha$+[N{\sc ii}]}

\documentclass{aa}
\usepackage{graphics}
\begin{document}

\title{Rings in the haloes of planetary nebulae\thanks{Based on
observations obtained at: the 2.5~INT telescope of the Isaac Newton
Group and the 2.6m~NOT telescope operated by NOTSA in the Spanish
Observatorio del Roque de Los Muchachos of the Instituto de Astrof\'\i
sica de Canarias; the 3.5m NTT and the 2.2 MPG/ESO at the European
Southern Observatory in Chile; and the NASA/ESA Hubble Space
Telescope, obtained at the Space Telescope Science Institute, which is
operated by AURA for NASA under contract NAS5-26555.}}

\author{R.L.M. Corradi\inst{1}, P. S\'anchez-Bl\'azquez\inst{2}, 
G. Mellema\inst{3}\thanks{\emph{Present address:} Netherlands
       Foundation for Research in Astronomy, P.O. Box 2, NL-7990 AA
       Dwingeloo, The Netherlands}, C. Gianmanco\inst{4}, \and
       H.E. Schwarz\inst{5}}

\institute{
	 Isaac Newton Group of Telescopes, Ap.\ de Correos 321,
	 38700 Sta. Cruz de la Palma, Spain
         \\e-mail: rcorradi@ing.iac.es
	\and 
         Departamento de  Astrof\'\i sica, Universidad Complutense, 
         28040, Madrid, Spain
         \\e-mail: pat@astrax.fis.ucm.es
	\and
        Sterrewacht Leiden, Postbus 9513, 2300 RA Leiden, The Netherlands    
          \\e-mail: mellema@strw.LeidenUniv.nl
	\and
        Instituto de Astrof\'\i sica de Canarias, 38200 La Laguna, 
        Tenerife, Spain
         \\e-mail: corrado@ll.iac.es
        \and  
            Cerro Tololo Inter-American Observatory,
            NOAO-AURA, Casilla 603, La Serena, Chile
           \\email: hschwarz@ctio.noao.edu
}
\offprints{R. Corradi}
\date{\today}

\abstract{
We present a search for rings or arcs in the haloes of planetary
nebulae (PNe). We discovered such structures in eight PNe, tripling
the sample of PNe with known rings. This shows that, contrary to what
was believed to date, the occurrence of mass loss fluctuations with
timescales of 10$^2$--10$^3$~yrs at the end of the asymptotic giant
branch phase (AGB) is common.  We estimate a lower limit of the
occurrence rate of rings in PN haloes to be $\sim$35\%.\\ Using these
new detections and the cases previously known, we discuss the
statistical properties of ring systems in PNe haloes. We estimate that
the mass modulation producing the rings takes place during the last
10000 or 20000~yrs of AGB evolution. In PNe, the spacing between rings
ranges from $<$0.01~pc to 0.06~pc, significantly larger than those
seen in proto-PNe. This, together with the finding of a possible
positive correlation of spacing with the post-AGB age of the nebulae,
suggests that the spacing of the rings increases with time.\\ These
properties, as well as the modest surface brightness amplitudes of
rings, are consistent with the predictions of the dust-driven wind
instability model explored by Meijerink et al. (2003), but do not
immediately exclude other proposed models.
\keywords{Planetary nebulae: general -- Stars: AGB and post--AGB }
}
\authorrunning{Corradi et al.}
\titlerunning{Rings in planetary nebulae haloes}
\maketitle   
 
\begin{table*}   
\caption{Log of the new observations.}
\begin{center}
\begin{tabular}{lclllc}    
\hline
Object           & PNG        & Telescope & Filter & \multicolumn{1}{c}{Exp. 
time} & Seeing    \\
                 &            &           &        & \multicolumn{1}{c}{[sec]} 
  & [arcsec]    \\
\hline		 	                                              
\object{NGC 40}  & 120.0+09.8 & INT       & \hanii & 60,3600 & 1.1 \\
\object{NGC 1535}& 206.4--40.5& INT       & \oiii  & 30,1200 & 1.5 \\
\object{NGC 3242}& 261.0+32.0 & INT       & \oiii  & 30,1800 & 1.6 \\
\object{NGC 6543}& 096.4+29.9 & NOT       & \oiii  & 30,120,1800 & 0.8 \\
\object{NGC 7009}& 037.7--34.5& INT       & \oiii  & 10,60,300,1200  & 1.4 \\
                 &            & INT       & \hanii & 20,180,1200     & 1.3 \\
                 &            & MPG/ESO   & \oiii  & 15,90,600    &  0.8   \\
\object{NGC 7027}& 084.9-03.4 & NOT       & \oiii  & 10,120,300,1800 & 0.8 \\
\object{NGC 7662}& 106.5--17.6& INT       & \oiii  & 20,110,600 & 0.8    \\
\end{tabular}
\end{center}
\label{T-obs}
\end{table*}

\section{Introduction}

Most PNe have multiple shells around their central stars. Modern
simulations allow us to interpret the formation of most of these
shells. We know, for instance, that the typical double-shell structure
of the bright inner body of round and elliptical PNe is the result of
the combination of wind interaction (producing the so-called inner
{\it rims}) and that photo-ionization effects are responsible for
producing the attached {\it shells}. For a detailed discussion see
e.g.~Mellema (\cite{M94}) and Sch\"onberner et al.\ (\cite{SSSKB}).

Around the inner nebula, with 1000 times lower surface brightness, an
extended ionized {\it halo} has been found in 60\% of the PNe for
which proper imaging has been obtained (Corradi et al. 2003, hereafter
CSSP03). These haloes are interpreted as being matter lost at the end
of the asymptotic giant branch (AGB) phase, their edges being the
signature of the last thermal pulse (Steffen \& Sch\"onberner
\cite{Ste03}).

In recent years, a new puzzling component has been discovered in the
inner regions of PNe haloes. High resolution imaging done mainly with
the Hubble Space Telescope has revealed the presence of so-called
`rings' in four PNe: Hb~5, NGC 6543, NGC 7027 (Terzian \& Hajian
\cite{TH00}), and NGC 3918 (CSSP03), as well as around six proto-PNe
and one AGB star (see the review by Su 2004).  The name `rings' is
somewhat misleading in that these structures just appear to be rings
when projected on the sky. They are more likely to be `shells'. But
since the nomenclature of morphological features in PNe is already
somewhat confused, we will use what seems to be the most widely
accepted name, namely `rings'. Note that Soker~(2002, 2004) refers to
them as `M-arcs'. Their formation, occurring when the star loses mass
at the highest rate during its evolution, is relevant to understanding
the physical processes producing the ultimate ejection of the envelope
of low- and intermediate-mass stars. To date, however, very little is
known about the physical and dynamical properties of these rings,
especially in the PNe phase where, so far, they were considered
to be a rare phenomenon.  In this paper, we present the results of an
extensive search for such rings in PNe haloes. We find them in eight
more PNe, showing that rings are quite common, and thus strengthening
the idea that the physical processes producing the rings is of
general importance to understand mass loss in the latest phase of the
AGB.

\section{Observations}

The observational targets were mainly chosen from the list of PNe with
haloes in CSSP03.  The original images were carefully reanalysed, and
for several targets for which hints of the existence of rings were
found we obtained new deep images.  It must be stressed that often the
main limitation in searching for this kind of structures is not the
spatial resolution of the observations, but the instrumental scattered
light as discussed in CSSP03. For this reason, we obtained most of the
new images with the prime-focus Wide Field Camera (WFC, pixel scale
0$''$.33) at the 2.5m~Isaac Newton Telescope (INT) at La Palma,
which has a clean point-spread function.  We obtained
deep \oiii\ and/or \hanii\ images.  The \oiii\ filter bandpass does
not include any other important emission lines beside the \oiii\
doublet at $\lambda$=$500.7$ and $495.9$~nm.  The
\hanii\ filter includes both the emission from hydrogen \ha\ and that
from the singly ionized nitrogen doublet at $\lambda$=$654.8$ and
$658.3$~nm.  Exposures were split into several sub-exposures to limit
the effects caused by over-saturation of the inner bright nebula (i.e.\
charge overflow).  For the same reason we sometimes positioned the
inner nebula in the gap between CCDs in the four-chip mosaic of the
WFC. \oiii\ images of NGC~7009 were also obtained with the Wide Field
Imager (WFI, pixel scale 0$''$.24) of the 2.2mMPG/ESO telescope
at La Silla, Chile, under good seeing conditions. \oiii\ images of
NGC~6543 and NGC~7027 were obtained with the 2.6m~Nordic Optical
Telescope (NOT) at the ORM and its multi-mode instrument ALFOSC (pixel
scale 0$''$.19).  The images were reduced in a standard way
using the IRAF and MIDAS packages. The observations with the INT+WFC
were partially reduced by R.~Greimel and C.~Davenport through the
instrument pipeline. A summary of the new observations is presented in
Tab.~\ref{T-obs}.

In addition, the HST archive was searched for the deepest images of a
number of nebulae known or suspected to have rings.

\begin{figure*} 
\resizebox{15.0cm}{!}{\includegraphics{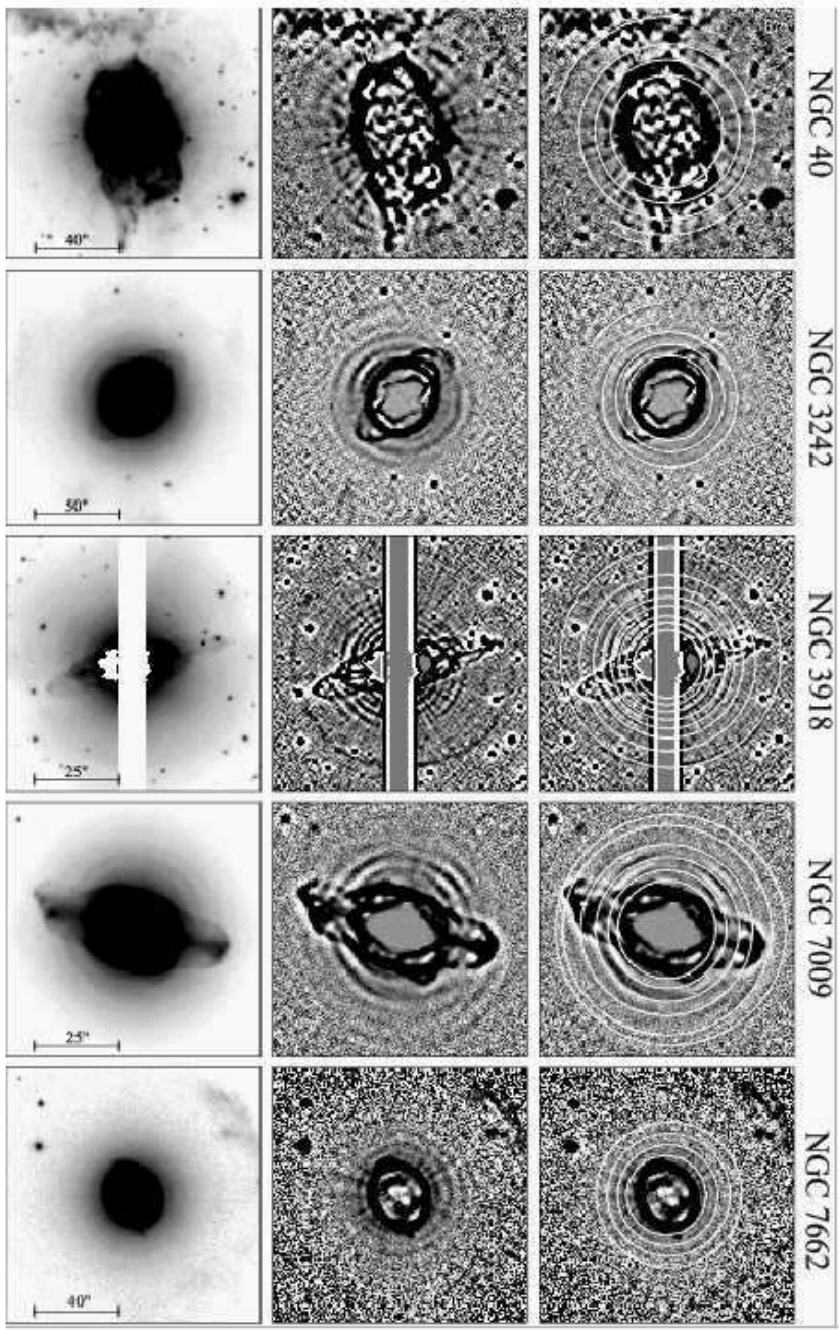}} 
\caption{Images of the definite new detections of rings. All images 
are displayed in {\it negative} greyscale (i.e. black means larger emission
values). On the left, the original image in a logarithmic display, at the
centre the {\it sda} processed image (see text) in a linear display, and to
the right the same ones with superimposed a visual circular fit of the
rings/arcs.}
\label{F-images1}
\end{figure*} 
\begin{figure*} 
\resizebox{15.0cm}{!}{\includegraphics{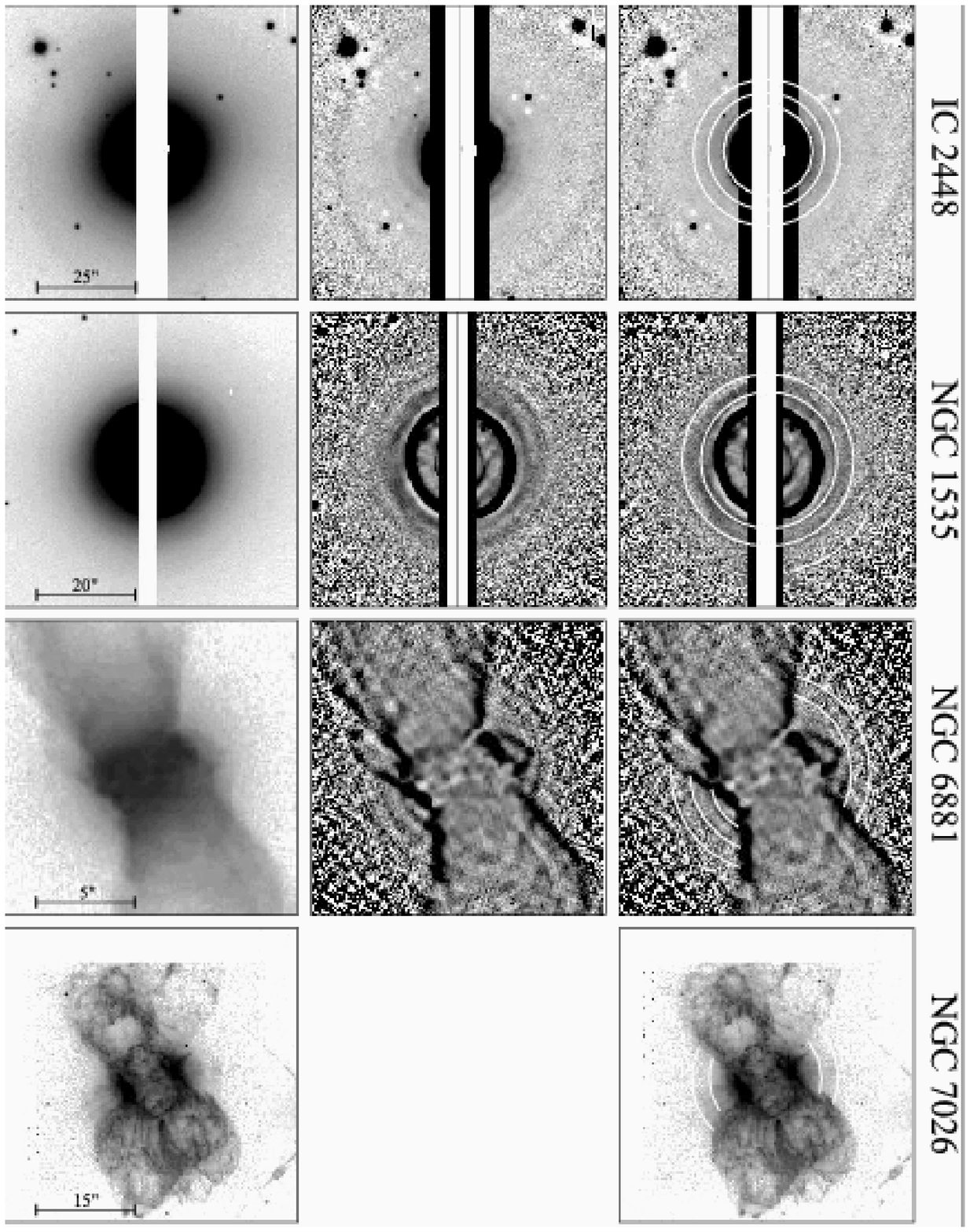}} 
\caption{As in Fig.~1, but for the PNe with probable detections of rings. 
For NGC~7026, no {\it sda} processed image is presented (see text).}
\label{F-images2}
\end{figure*} 

\section{Image analysis: finding and enhancing the rings}

The new images clearly reveal the existence of rings in several
of our target PNe. However, rings are not easily visualized in
greyscale or colour plots, as they are located in the inner regions of
the haloes whose surface brightness has a very steep radial profile
with a large dynamic range. In order to better highlight the
rings and measure their properties, we processed the images in
several different ways. Good results are obtained by taking the
logarithmic derivative of the images. This method is described in
Corradi et al. (2004). An even better way to enhance the rings is to
divide the original image by its smoothed version, using any kind of
algorithm with a smoothing scale-length of the order of the rings
spacing. We call this as the {\it smoothing} algorithm. Nearly
identical results are obtained by processing images with what we call
the {\it sda} ({\it shift}, {\it divide}, and {\it add}) algorithm.
One first produces four ``shift'' images, by applying to the original
frame $I(x,y)$ a shift of $p$ pixels to the right, left, up and down,
respectively. Then the original image is divided by each of the shift
images, and the four ratio maps are summed up together producing the
final image $I_{sda}$.  In symbols:
\begin{eqnarray}
\nonumber
I_{sda}=\frac{I(x,y)}{I(x-p,y)}\hspace*{-1mm}+\hspace*{-1mm}
\frac{I(x,y)}{I(x+p,y)}\hspace*{-1mm}+\hspace*{-1mm}
\frac{I(x,y)}{I(x,y-p)}\hspace*{-1mm}+\hspace*{-1mm}
\frac{I(x,y)}{I(x,y+p)}\, .
\end{eqnarray}
In this way, pixels corresponding to the larger surface brightness of
the rings are enhanced by division by an ``average'' surface
brightness of the halo in that region.  The {\it sda} procedure
was tested in several ways, to make sure that no artificial rings are
created by the algorithm and that the location of the rings remains
the same.  We created a model image with a $r^{-3}$ surface brightness
profile and, superimposed, sinusoidal fluctuations with amplitude
20\%\ the value of the local intensity and period equal to 20 pixels,
simulating a halo with rings.  The {\it sda} algorithm was then
applied using shifts $p=10,15,20,25,30$~pix.  For any choice of $p$,
rings are effectively enhanced.  The only spurious effect is the
appearance, for shift values $\ge 20$~pix, of slight distortions of
the circular symmetry, but even in the directions where the effect is
the largest, the original spacing between rings is always preserved
with a high accuracy.


We processed with the three algorithms all the new ground-based
images, as well as the \oiii\ images of IC~2448, NGC~1535 and NGC~3918
from CSSP03, and the HST images of NGC~6881 (\ha) and NGC~7026
(\nii). For the {\it sda} processing, according to the results of the
tests, a shift value equal to or smaller than the average ring spacing
was adopted. In all PNe, rings show up using any of the three
algorithms (and except for IC~2448 and NGC~1535 they are also clearly
detected in the non-processed images). Faint rings are better enhanced
using the smoothing algorithm or the {\it sda} one, rather than taking
derivative images.  The {\it sda} algorithm, which produces nearly
identical results to the smoothing method for the external regions,
provides a better enhancement of the innermost ring, as smoothing in
this region is affected by the abrupt change of slope of the surface
brightness profile due to the nearby bright rim and shell.  

Logarithmic greyscale displays of the original frames, the images
processed using the {\it sda} algorithm, and visual fits of the
detected rings are presented in Figs.~\ref{F-images1} and
~\ref{F-images2}.

\begin{table*}   
\caption{Rings properties in PNe. In the second column, $r$ means ``rings'' 
and $a$ ``arcs''. In the last column, we quote the adopted distance
for each PN.}
\begin{tabular}{lllclc}    
\hline
Object& N.of rings& Spacing   & \multicolumn{1}{c}{Peak-to-continuum} & 
Comments & Dist.\\
      &           & [arcsec]  & \multicolumn{1}{c}{ratio}     &              & 
[kpc] \\
\hline	     				           
\multicolumn{6}{l}{\phantom{pi} \it Clear detections}\\
\object{Hb 5}$^\star$ & 6$a$    & 0.6-1.3   &  2   &  concentric               
   & 1.5 \\
\object{NGC 40}     & 3$r$+2$a$ & 5.0-7.4   & 1.10 & concentric                
   & 1.1 \\
\object{NGC 3242}   & 3$r$+2$a$ & 6-10      & 1.12 & concentric                
   & 0.8 \\
\object{NGC 3918}   & 8$r$      & 2.0-6.6   & 1.07 & concentric (except last)  
   & 1.2 \\
\object{NGC 6543}$^\star$ &$>$11$r$&2.2-3.8 & 1.2  & concentric                
   & 1.0 \\
\object{NGC 7009}   & 6$r$      & 2.9-4.8   & 1.09 & non-circular, intersect?  
   & 0.9 \\
\object{NGC 7027}$^\star$ & $>$9$r$&2.2-5.8 & 1.5  & incomplete,some 
intersecting & 1.0 \\
\object{NGC 7662}   & 4$r$      & 5.0-5.3   & 1.12 & approx concentric         
   & 1.2 \\[2pt]
\multicolumn{6}{l}{\phantom{pi} \it Probable detections}\\
\object{IC 2448}    & 3$r$      & 3.5-3.8   & 1.05 &  non circular             
   & 1.4 \\
\object{NGC 1535}   & 2$r$+1$a$ & 7.2-11.3  & 1.07 & non-circular & 1.8 \\
\object{NGC 6881}   &  3$a$     &   1.0     &  --  &            & 3.2 \\
\object{NGC 7026}   &  2$a$     &   2.5     &  --  &            & 1.6 \\
\end{tabular}
\\[5pt]\indent $^\star$Data from Su (2004).
\label{T-ringprop}
\end{table*}

\section{Description of individual nebulae}

The individual PNe are discussed below, and a summary of the properties
of their ring systems (including objects discussed by Su 2004) is
presented in Tab.~\ref{T-ringprop}.

\subsection{IC 2448} 
CSSP03 conservatively put this object in the list of PNe with no haloes
because of the possibility that the diffuse, featureless luminosity detected
around the central body of the PN is instrumental scattered light. The
processing of their original, high-quality \oiii\ image (seeing 0$''$.6)
reveals the presence of fluctuations of the radial surface brightness profile
that can be roughly described as a system of three or more rings in an
extended AGB halo. Rings are not as well defined as in other PNe, and show
some sign of non-circularity and/or offsets from the central star.

\subsection{NGC 40}
In addition to the structure in the faint halo around this
nebula, our deep \hanii\ image reveals the existence of a system of
three concentric inner rings centred on the central star.
Fragments of one or two external rings are also found.  NGC 40 is a
low excitation nebula, and no halo is visible in the \oiii\ images.

\subsection{NGC 1535}
Processing of the original \oiii\ image by CSSP03 (seeing 0$''$.6)
reveals the presence of at least two rings and a fragment (arc) of an
outer one. Rings are broad and non circular. A recent image taken with
the INT+WFC, in spite of having a lower spatial resolution (seeing
1$''$.5), confirms the presence and properties of such broad rings,
thus excluding the possibility that they are instrumental artifacts.

\subsection{NGC 3242}
The \oiii\ INT+WFC image shows three clear rings and fragments of two
other outer ones. The innermost two rings are faintly visible in
archival HST images. The rings are concentric and the spacing between them
is variable. They are also visible in the \hanii\ image by CSSP03.

\subsection{NGC 3918}
The rings in the halo of this nebula were noted by CSSP03. Data
processing of their image confirms that there are at least 8
concentric circular rings with a spacing that increases from 2$''$ for the
innermost ones, to 3$''$.8 for the second last one. The outermost
ring is slightly offset by a couple of arcseconds to the North-West
and has a radius 6$''$.6 larger than the preceding one (but we might
be missing an intermediate ring). It also appears to be more intense
than the previous few ones, defining a sharp edge of the system of
rings. The actual edge of the AGB halo, presumably corresponding to
the last thermal pulse, is at a much larger distance from the central
star, see CSSP03.

\subsection{NGC 6543}
Our deep \oiii\ images confirm the results of the detailed analysis of
the HST data by Balick et al. (2001), with the possible addition of
two more outer rings out to a distance of $\sim 50''$ from the central
star. For this reason, our new images are not presented in
Figs.~\ref{F-images1} and \ref{F-images2}.

\subsection{NGC 6881}
Three roughly circular, equally spaced arcs are barely visible in \ha\
HST archive images along the directions perpendicular to the bipolar
lobes.

\subsection{NGC 7009}
The new INT and MPG ground-based images reveal the existence of at
least six rings in the inner regions of the knotty halo of this PN. As
in the other cases, rings are best visible in the {\it sda} processed
image, and show clear deviations from the circular symmetry (in
Fig.~\ref{F-images1}, a visual fit for the northern side of the
nebula fits the southern side poorly), with possible intersection
between adjacent rings. The spacing between the rings is variable.

\subsection{NGC 7026}
Two arcs (whose centre is slightly offset from the central star) are
possibly identified in \nii\ HST images, but this is the weakest case
of all our new detections. This is in fact the only case in which our
image processing does not enhance the possible rings, probably because
of the highly structured local surface brightness distribution
(including strong radial features).

\subsection{NGC 7027}
The reflection rings of this object were discussed by e.g. Su (2004).
Our new deep \oiii\ images (not presented in Figs.~\ref{F-images1} and
\ref{F-images2}) confirm their results, showing evidence for a few other
faint outer rings up to a distance from the central star of $\sim50''$.

\subsection{NGC 7662}
A new, clear system of 4 rings is found in this PN. The rings are
approximately concentric, but with some hints of non-circularity and
offset from the central star (the latter for the outer rings).

\section{General properties of rings}

\subsection{Detection rate} 

Rings are found in PNe of different morphological classes: elliptical
nebulae (NGC~40, NGC~1535 and IC~2448), ellipticals with
low-ionization small-scale structures like FLIERS (NGC 3242 and NGC
7662), more collimated nebulae with jets (NGC~3918, NGC~6543 and NGC
7009), and bipolars (Hb~5, NGC~6881, NGC~7026, and NGC~7027, the latter
also probably bipolar, see Bains et al. 2003).  Note also that some of
the PNe with detected rings, namely NGC~40, NGC~6543 and NGC~7026 have
hydrogen deficient central stars.

CSSP03 list 21 PNe possessing bona-fide AGB haloes. Including the
haloes with rings revealed in this paper for IC~2448, NGC~40, and
NGC~1535, this makes 24 non-bipolar PNe known to have AGB haloes. The
present analysis of the CSSPO3 sample (including the new deep images
and the HST archive ones) allowed us to discovered rings in 8 of these
24 PNe, i.e. some 35\%\ of the sample of AGB haloes
investigated. Considering that CSSP03 estimate that $\ge$60\% of the
whole sample of Galactic PNe have ionized AGB haloes, we derive that
the lower limit for the frequency of rings in round and elliptical PNe
is 20\%.
The real figure can be much higher, as high-quality images needed to
detect rings (or even to reveal haloes) are not available for the
whole CSSP03 sample.  One is therefore tempted to ask whether rings
exist in {\it all} PNe?  The database of CSSP03 contains some haloes
with no apparent evidence for rings.  These are for example CN~1-5, IC
2165, NGC 2022, NGC 2792, NGC 6826 and PB4.  Except for NGC 6826,
however, all these PNe are located at systematically larger distances
(several kpc) than those with detected rings, most of which lie at
about 1~kpc from the Sun, see Tab.~\ref{T-ringprop}. This is also
reflected by a systematically smaller apparent size of the haloes of
PNe with no rings. The resolution in the ground-based images just may
not be sufficient to detect rings in those distant PNe.  We therefore
conclude that the 35\% occurrence rate in the present study is a lower
limit, and the possibility that rings are present in the majority of
or all PNe cannot be excluded.

The situation for bipolar PNe is more uncertain, as we do not even
know whether this class of objects has AGB haloes, at least in the
sense as defined by CSSP03. However, having found arcs in several
objects, it is likely that also for this class of nebulae circular
rings/arcs are common. Therefore bipolar lobes would excavate through
pretty spherical initial circumstellar density distributions,
exacerbating the problem of a sudden turn from spherical to highly
collimated mass loss at the end of the AGB (e.g. Sahai \cite{Sah02}).
This is supported by the finding of rings and arcs in several {\it
bipolar} proto-PNe (Su 2004).  With the present data, we cannot say if
the rings of Hb~5, NGC~6881 and NGC~7026 are seen in direct or
reflected light as is the case in NGC~7027 and the proto-PNe.

\subsection{Post-AGB and post-ionization ages} 

Some of the models proposed to explain the formation of rings predict that
during the PN phase rings would persist only few thousand years after
photo-ionization (Meijerink et al. 2003, hereafter MMS03).  Unfortunately,
measuring how long the ring systems have been photo-ionized is a hard task
for individual PNe. In principle, this could be done if the luminosity and
temperature of their central stars (CSs) are accurately known. A comparison
with the theoretical tracks would then allow us to determine their stellar
mass and post-AGB age.  This is however extremely difficult to do, as the
evolutionary timescales are strongly dependent on mass, which on the
horizontal part of the post-AGB tracks mainly depends on luminosity, the
determination of which suffers from the large uncertainties in the distance
to individual PNe. We attempted this kind of analysis, using both the data
from the CSs in CSSP03 and those from Mal'kov (1997), but for individual
objects the results show large discrepancies between the two different data
sets, making this kind of analysis impracticable.  All we can conclude at
present, in a statistical sense, is that PNe with rings do not exhibit a
clear trend in their locations in the HR diagram as compared to the global
sample of PNe haloes as shown in Fig.~4 of CSSP03.  In addition, even if the
CS could be precisely located in the HR diagram, the post-AGB age at which
haloes and their rings became photo-ionized cannot be accurately determined,
as it depends not only on the CS mass but also on the circumstellar density
distribution at the end of the AGB, which is not well known.  More details
about this point can be found in Perinotto et al. (2003), who showed that the
post-AGB age at which the inner halo becomes photo-ionized is $\sim$3500 yrs
for a `standard' CS of M=0.605~M$_\odot$ and the AGB mass distribution
computed by Steffen
\& Sch\"onberner (\cite{Ste03}). The situation is more uncertain for
larger masses (M$\ge$0.625~M$_\odot$) for which this time scale
would be lower although it may also happen that these haloes are never
ionized.

\begin{figure} 
\resizebox{\hsize}{!}{\includegraphics{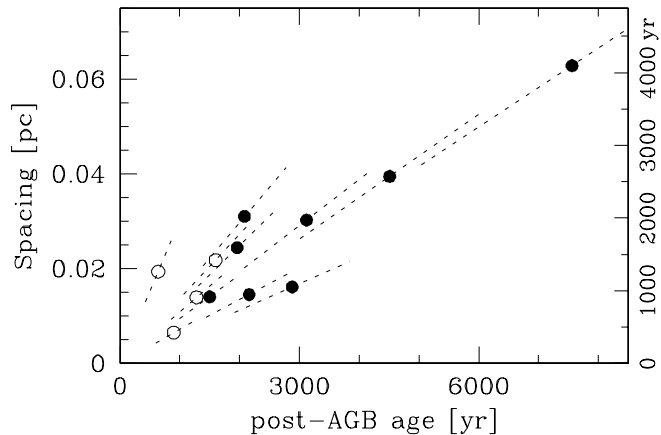}} 
\caption{Average spacing vs. the kinematic post-AGB age 
for elliptical PNe (filled circles) and bipolars (open
circles).}
\label{F-spa_age}
\end{figure}

An alternative measurement for the age of a PN is the kinematic age:
its size divided by its expansion velocity. According to the models in
Sch\"onberner et al. (1997), the kinematic age of the attached shell
of an elliptical PN provides a better estimate of the true post-AGB
age of the nebula than that of the rim. We list on the abscissae of
Fig~\ref{F-spa_age} the kinematic ages of the elliptical PNe with
rings, computed for their shells using the kinematic data from the
literature or from our unpublished echellograms (except for NGC 40
where no attached shell is present, we used the age of the rim
instead). To compute the linear sizes, we used the distances listed in
the last column of Tab.~\ref{T-ringprop}, which were determined by
taking a weighted average of the values from the catalogue of Acker et
al.~(\cite{SESO}), or taken from recent papers with individual
distance determinations (Reed et al. \cite{Re99} for NGC 6543 and
Palen et al. \cite{Pa02} for IC~2448).  All nebulae except NGC~1535
(age=7500~yrs) have kinematic post-AGB ages smaller than 4500~yrs. If
we subtract from this age the time from the end of the AGB to the
first photo-ionization of the halo (3500~yrs for a CS of
0.605~M$_\odot$\footnote{Obviously, for post-AGB ages smaller than
3500~yrs, the photo-ionization of the halo must have occurred
earlier.}), the numbers imply that, in a broad statistical sense, the
observed rings are still in the phase before they are predicted to
vanish because of photo-ionization effects (MMS03). This conclusion
would still hold if a slightly different distance scale, like the one
in CSSP03, is adopted.

The post-AGB ages of bipolar PNe in our sample are not relevant in
this discussion because their rings may not be ionized; in any case,
they are all younger than two thousand years according to the adopted
distances and spatiokinematic studies in the literature (Solf \&
Weinberger
\cite{SW84}, Corradi \& Schwarz
\cite{CS93}, Guerrero \& Manchado \cite{Gue98}, Bains et
al. \cite{Ba03}).

\subsection{Duration of the mass loss modulation producing the rings}

Once the distance and expansion velocity of a PN are known, the radius of the
outermost ring allows us to estimate the age of the system.  We assume a halo
expansion velocity of 15~\kms\ (see CSSP03) for all PNe, since individual
kinematic measurements are mostly lacking, a general handicap when studying
PN haloes.

\begin{figure}
\resizebox{\hsize}{!}{\includegraphics{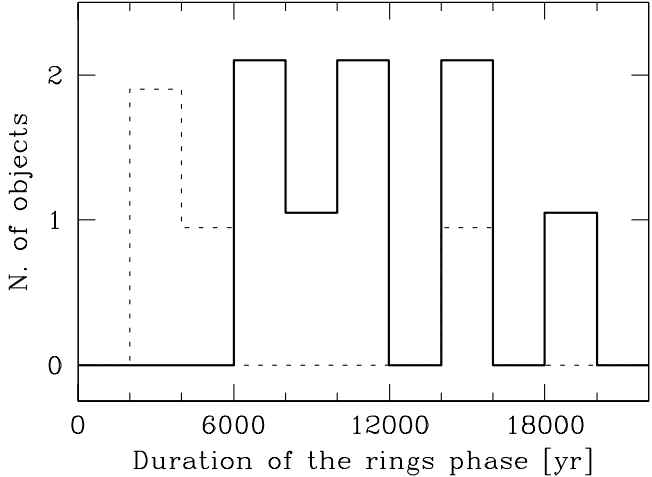}} 
\caption{Duration of the ``rings'' phase for elliptical PNe 
(solid histogram line) and bipolars (dotted line).}
\label{F-totalages}
\end{figure}

If we subtract the ``kinematic'' post-AGB age of the PNe from the age of the
outermost ring, we obtain an estimate of the duration of the phase of mass
loss modulation at the end of the AGB producing the rings.  This is shown in
Fig.~\ref{F-totalages}: the duration of the phase spans from 6000~yrs for
IC~2448 and NGC~7009 to a maximum of 14000-18000~yrs for NGC~40, NGC~1535,
NGC~6543 and NGC~7027.  The other bipolar PNe show durations less than 5000
yrs, but they should be considered highly uncertain because the region where
rings are found is very faint, and we may have missed some rings.

Therefore, since it is likely that fainter rings in other PNe in our
sample were missed, we conclude that the phase of mass modulation
giving rise to the rings might characterize the last 10000 to
20000~yrs of AGB evolution. CSSP03 showed in their Tab.~4 that for
these PNe the time elapsed since the last AGB thermal pulse (whose
signature is the limb-brightened edge of the AGB halo) is typically
factors 1.8 to 4 longer than the ages in Fig.~\ref{F-totalages}.
Thus, even when we missed a few faint rings, we conclude that the
formation of rings is not associated with thermal pulses.

\subsection{Ring geometry and spacing} 

All PNe of our sample show some variation in the spacing between
the rings. The most regular cases seem to be NGC~40 and NGC~7662. Most other
systems of rings (Hb~5, NGC~1535, NGC~3242, NGC~3918, NGC~6543, NGC~7009)
have definitely a variable spacing, with some (weak) evidence that
spacing increases with distance from the central star. NGC~1535, NGC~7009 and
NGC 7027 also show clear deviations from circularity, with some rings
intersecting each other and in some cases with centres offset from the central
star. IC~2448 and NGC~7026 have poorly defined rings and are not considered
in this analysis.

The range of spacing values spanned by each system of rings is listed in
Tab.~\ref{T-ringprop}. The average spacing for each PN, transformed to a
linear scale using the adopted distances, is displayed in the ordinate of
Fig.~\ref{F-spa_age}, and ranges from 0.007~pc (Hb~5) to 0.063~pc
(NGC~1535). There is also an apparent trend for bipolar PNe (dotted
histogram) to have smaller spacings than elliptical PNe, but this is not
statistically significant because of the small sample.  Adopting an expansion
velocity of 15~\kms, these spacings corresponds to timescales for the mass
loss fluctuations of between 400 and 4000~yr (see the right y-axis of
Fig.~\ref{F-spa_age}).

Most interestingly, Fig.~\ref{F-spa_age} shows that the average spacing
correlates with the post-AGB age of the nebula, suggesting the existence of
evolutionary effects in the ring spacing. This is even more clearly indicated
by the fact that the spacing in the six proto-PNe listed by Su (2004) is
systematically and significantly smaller than that of our sample of PNe,
ranging between 80 and 600~yr if the same expansion velocity of 15~\kms\ is
adopted. In Fig.~\ref{F-spa_age}, we indicate with dotted lines how points
would move if the distances were off by a factor of two (33\% on each side of
the adopted value).  The distance enters linearly in both the estimate of the
linear spacing and the kinematical age and may therefore cause a false
correlation to appear. From the dotted lines it can be seen that if the
distances of the three objects (NGC~1535, NGC~40 and NGC~7662) that better
define the linear correlation in Fig.~\ref{F-spa_age}, are overestimated by a
factor of $\mincir$$2$, the evidence for the existence of the correlation
would be weaker. Our result should therefore be treated with some caution.

\subsection{Surface brightness profiles and peak-to-continuum ratio}

Radial surface brightness profiles of the inner haloes of our sample
of elliptical PNe were extracted at position angles corresponding to
the minor axis of the rims/shells. In most cases, a power law
$I\propto r^{-\gamma}$ provides a fairly good fit to the surface
brightness profiles in the region of the rings (especially for
IC~2448, NGC~3242, and NGC 7662), with $\gamma$ ranging from 3.3 to
4.5. In the case of IC~2448 and NGC~3242, $\gamma$ decreases abruptly
to shallower slopes at the end of the region occupied by the rings; in
other cases, like NGC~40, NGC~1535, NGC~3918 and NGC~7009, the slope
of the power-law shows a more systematic and continuous decrease with
radius, and this is even more pronounced in the two PNe with no rings
(NGC~6826 and NGC~2022) which we consider for comparison.  These
surface brightness slopes correspond to density profiles in the
regions of the rings steeper than $\rho^{-2}$, in some cases as steep
as $\rho^{-3}$. This confirms that mass loss increases substantially
at the end of the AGB phase (e.g.~Steffen \& Sch\"onberner
\cite{Ste03}).

Once this large-scale trend was removed, we estimated the surface brightness
contrast between the ring peaks and the (pseudo)continuum (or, in other
words, half the contrast between the ring peaks and dips). Limited spatial
resolution (through seeing or finite mirror size) reduces the true
peak-to-continuum intensity contrast. An estimate of this can be found by
comparing ground-based and HST data. We compared the HST ACS \oiii\ images of
NGC~6543 (instrumental resolution 0$''$.05) with those obtained at the NOT
(seeing 0$''$.8). For different rings (whose typical width is 2$''$ and
spacing 2$''$.9), the peak-to-continuum ratio in the NOT images is measured
to be between 5\% and 10\% lower than in the HST images, where it reaches a
maximum value of 1.2. In other PNe of our sample, spacing between rings is
equal to or larger than that of NGC 6543, so that if the ratio between the
width and spacing does not vary dramatically, we do not expect that the
estimate of the peak-to-continuum ratios using our ground-based images
underestimates the real values much more than the amount found for NGC~6543.

In all nebulae, we find that the peak-to-continuum intensity ratio is
strongly variable within the same system of rings. The maximum
peak-to-continuum ratio measured for each PN is reported in
Tab.~\ref{T-ringprop}.  Even considering seeing effects, these numbers are
systematically lower than for the sample of proto-PNe in Su (2004). We find
however, no (anti)correlation with the kinematical post-AGB age of our
nebulae. This could be explained by the large dispersion and irregularity in
the peak-to-continuum contrast for individual rings within the same nebula,
which make the values listed in Tab.~\ref{T-ringprop} just an
order of magnitude number. Note also that the rings in proto-PNe show up in
reflected light, whereas the ionized haloes produce intrinsic emission, which
may also account for some of the differences.

For bipolar PNe, reliable measurements of the peak-to-continuum
surface brightness can only be done for Hb~5 and NGC~7027. The values
listed by Su (2004) are also listed in Tab.~\ref{T-ringprop}.

\section{Discussion, interpretation and conclusions}

In spite of the paucity of data available so far, the origin of rings
in PNe haloes has been widely debated, and a number of formation
mechanisms has been proposed. These include binary interaction
(Mastrodemos \& Morris 1999), magnetic activity cycles (Soker 2000,
Garc\'\i{a}-Segura et al.~2001), instabilities in dust-driven winds
(Simis et al.~2001) and stellar oscillations (van Horn et al.~2003,
Zijlstra et al. 2002, Soker 2004). These models all produce rings
morphologically similar to the observed ones, and only a detailed
study of their physical and dynamical properties will allow us to
distiguish the models. A first start of this was made by
MMS03, who followed the evolution of rings produced by dust-driven
wind instabilities using radiative hydrodynamic modelling, resulting
in emission properties, line ratios, and line shapes.

The main differences between the models are in the way the rings are
supported, and, at least in the published versions of the models, the
amplitudes of the variations. In the case of magnetic field reversals,
the rings are supported by magnetic fields, material collecting in
areas of low magnetic pressure. This makes for stationary rings and
the authors claim that this will provide for the required longevity.
In the other cases the rings are dynamic, corresponding to waves
(density and velocity variations) set up by variations in the mass
loss either caused by variations at the base of the wind (dust-driven
wind instabilities or stellar oscillations), or created by the
interaction with a companion star. Such dynamic rings will ultimately
disappear, the time scale depending on the amplitude of the waves, and
the sound speed in the wind material. The latter means that this
process will be slow while the wind is neutral and speed up after
ionization. A simple estimate can be derived from the ratio of the
extent of the ring area and the sound speed, giving values of
100\,000~yrs during the neutral phase and several 1000~yrs during the
ionized phase.

The distinction between magnetic and dynamically supported rings is
not as clear cut as that though, since photo-ionization will raise the
thermal pressure of the wind, making the magnetic pressure
insignificant compared to the thermal pressure. This probably means
that the evolution of the rings during the ionized phase will be
similar in all models, but detailed magneto-hydrodynamic modelling
should confirm this.

The simulations of MMS03 showed that the original velocity variations
in the wind in case of dynamically supported winds is only important
for survival of the rings during their neutral phase. Raising the
pressure of the rings by photo-ionization will strengthen the waves
and much higher velocity variations are then induced.

The simulations in MMS03 show a number of properties which are
consistent with our new sample of rings. Firstly, the spacing of the
rings tends to increase with time. This is due to the slow merging of
the waves, a process which takes place from the start in the
dynamically supported case, and will probably only start after
ionization in the magnetically supported case. Although the
indications for this in the data are not conclusive, comparing the
typical spacing in the proto-PNe from Su (2004) with our PNe shows
this to be significant. The models also predict a weakening of the
amplitude of the rings over time, for which we also find some, albeit
marginal, evidence.

As to the observed amplitudes, one should realize that the rings are
not discrete entities, but rather density variations which we see in
projection. It is therefore not trivial to interpret the observed
values which may also be affected by temperature variations, boosting
the forbidden lines of \oiii\ and \nii. Taking the published results
of MMS03 we see that the amplitudes are rather modest, of order 1.1
(see their Figs.~10 and 11) and therefore comparable to the observed
values. The slope of the surface brightness profile seems to be
shallower than in the observations, but since the average AGB mass
loss was taken to be constant in the simulations, this is not
surprising.

In general then, the observed properties of the rings make sense
within the frame work of the dust-driven wind instability model
explored by MMS03. However, as these authors pointed out, this does
not necessarily mean that other models are excluded. Basically, any
model which leads to dynamically supported rings can be expected to
give such results, although it would be nice to see this confirmed by
simulations, especially for the magnetic reversal models.  In any
case, all models should now face the evidence that rings are
frequently found (and in all morphological classes), and the proposed
formation mechanism should apply to a large fraction of AGB stars
producing PNe.  

Concluding, we have shown in the present paper that rings are common
in PNe haloes, and thus of general relevance in the discussion of the
large mass loss increase that characterises the latest AGB
evolution. Testable predictions (especially in terms of the physical
and dynamical properties of the rings) which are able to distinguish
among all the different formation mechanisms proposed are however
presently lacking, and would be highly desirable in order to allow
further progress in this recent and important issue of the AGB and
post-AGB evolution.

\begin{acknowledgements}

We thank M. Azzaro and R. Greimel at the ING for taking some of the
images at the INT during service time, and R. Mendez and I. Saviane at
ESO for taking the WFI images of NGC~7009.
The research of GM has been made possible by a fellowship of the Royal 
Netherlands Academy of Arts and Sciences. 

\end{acknowledgements}

\end{document}